\documentclass[useAMS]{mn2e}
\usepackage{times}
\usepackage{graphicx}
\def\bbbr{{\rm I\!R}}

\begin{document}
\title{Eclipsing binaries in extrasolar planet transit surveys: 
 the case of SuperWASP}
\author[Willems, Kolb \& Justham]
{B. Willems$^{1}$\thanks{E-mail: b-willems@northwestern.edu, 
    u.c.kolb@open.ac.uk, sjustham@astro.ox.ac.uk }, U. Kolb$^{2,\star}$ 
    and S. Justham$^{2,3,\star}$ \\ 
    $^1$Northwestern University, Department of Physics and Astronomy, 
    2145 Sheridan Road, Evanston, IL 60208, USA \\
    $^2$Department of Physics and Astronomy, The Open University,
    Walton Hall, Milton Keynes, MK7 6AA, UK \\
    $^3$Department of Astrophysics, Oxford University, Oxford OX1 3RH, UK}

\date{Accepted ... Received ...; in original form ...}

\pagerange{\pageref{firstpage}--\pageref{lastpage}} \pubyear{2005}

\maketitle

\label{firstpage}

\begin{abstract}
Extrasolar planet transit surveys will also detect eclipsing
binaries. Using a comprehensive binary population synthesis scheme, we
investigate the statistical properties of a sample of eclipsing
binaries that is detectable by an idealised extrasolar planet transit
survey with specifications broadly similar to those of the SuperWASP
(Wide Angle Search for Planets) project.

In this idealised survey the total number of detectable single stars
in the Galactic disc is of the order of $10^6-10^7$, while, for a
flat initial mass ratio distribution, the total number of detectable
eclipsing binaries is of the order of $10^4-10^5$. The majority
of the population of detectable single stars is made up of
main-sequence stars ($\approx 60$\%), horizontal-branch stars ($\approx
20\%$), and giant-branch stars ($\approx 10\%$). The largest
contributions to the population of detectable eclipsing binaries stem
from detached double main-sequence star binaries ($\approx 60\%$),
detached giant-branch main-sequence star binaries ($\approx 20\%$), and
detached horizontal-branch main-sequence star binaries ($\approx
10\%$). White dwarf main-sequence star binaries make up approximately
0.3\% of the sample. 

The ratio of the number of eclipsing binaries to the number of single
stars detectable by the idealised SuperWASP survey varies by less than
a factor of 2.5 across the sky, and decreases with increasing Galactic
latitude. It is found to be largest in the direction of the Galactic
longitude $l=-7.5^\circ$ and the Galactic latitude $b=-22.5^\circ$.

We also show that the fractions of systems in different subgroups
of eclipsing binaries are sensitive to the adopted initial mass ratio
or initial secondary mass distribution, which is one of the poorest
constrained input parameters in present-day binary population
synthesis calculations. This suggests that once statistically
meaningful results from transit surveys are available, they will be
able to significantly improve the predictive power of population synthesis 
studies of interacting binaries and related objects.
\end{abstract}


\section{Introduction}

An ever increasing number of extrasolar planets is being discovered
(to date the number stands at more than 160), mostly
through the detection of the reflex motion of their parent star around
the planetary system's centre of mass. Planets with suitably oriented
orbits in space may also reveal their presence when they
transit the disc of their parent star and the transit is accompanied
by a measurable decrease in the observed stellar flux. When combined
with high-precision radial-velocity measurements, the transits yield a
wealth of information on the planet, such as its mass, radius, and
mean density.

The first successful planetary transit detections were made
independently by Charbonneau et al. (2000) and Henry et al. (2000),
who observed a clear drop in the lightcurve of the star HD\,209458 at
times consistent with the ephemeris of its already known
planetary-mass companion.  Since then, the search for transits has
rapidly evolved into a fully-fledged and widely used technique to hunt
for extrasolar planets. 

In anticipation of upcoming space-based exoplanet transit surveys such
as COROT (scheduled to launch in 2006) and Kepler (scheduled to launch
in 2007), ground-based surveys can pave the way towards obtaining a
statistically significant sample of planets in a relatively short
amount of time. Such a sample will be of great importance in improving
our understanding of the formation and evolution of extrasolar planets
as well as for detecting possible correlations between the presence of
planets and the properties of their host stars.
For an overview of some existing exoplanet
transit surveys, we refer to the reviews by Horne (2002, 2003) and
Charbonneau (2003). Despite the multitude of exoplanet transit
surveys, so far only six confirmed planets have been discovered by
direct observations of transits: five by the Optical Gravitational
Lensing Experiment (OGLE) (Konacki et al. 2003; Konacki et al. 2004;
Bouchy et al. 2004; Pont et al. 2004; Konacki et al. 2005), and one by
the Trans-Atlantic Exoplanet Survey (TrES) network (Alonso et
al. 2004).

Besides planetary transit detections, wide-field photometric transit
searches will also yield a wealth of photometric data on all kinds of
new and existing variable stars, including eclipsing binaries. Brown
(2003) estimated the rate of false planetary detections due to
eclipsing binaries with main-sequence and giant-type component stars
to be almost an order of magnitude larger than the rate of true
planetary detections.  Examples of astrophysical false positives in
wide-field transit searches are grazing incidence eclipses and the
blending of the light of an eclipsing binary with the light of a third
star (see Charbonneau et al. 2003 for details).

Our aim in this exploratory paper is to perform a binary population
synthesis study to predict the statistical properties and the Galactic
distribution of eclipsing binaries detectable by an extrasolar planet
transit survey, and to investigate their dependency on the initial
mass ratio distribution of the component stars.  Specifically, we
study the detectable sample for an idealised survey that is broadly
representative of the ground-based SuperWASP Wide Angle Search for
Planets (Kane et al. 2003, 2004; Street et al. 2003, Christian et
al. 2004, Pollacco 2005). The SuperWASP setup currently consists of 5
ultra-wide field lens cameras backed by high-quality CCDs, but is
designed to hold up to a total of 8 such cameras and CCDs (the upgrade
to 8 cameras is expected to be completed in 2005). The photometric
precision of the CCDs is largest in the 7--13\,mag magnitude range,
where it is aimed to be better than 10\,mmag (see Kane et al. 2004 for
a detailed discussion of the photometric accuracy of the prototype
WASP0).  Each camera on the SuperWASP setup furthermore has an
approximate field of view of $8\degr \times 8\degr$, so that,
accounting for some overlap, the total area covered by 4 cameras
operating simultaneously in a rectangular configuration is about
$15\degr \times 15\degr$. The total field of view for the full
8-camera setup will be $30\degr \times 15\degr$. The large field of
view is expected to yield precise photometry of approximately 
5\,000 to 10\,000 stars at a time per operational camera, and thus
40\,000 to 80\,000 stars at a time for 8 cameras operating
simultaneously. 

In the following sections, we study in some detail the relative
numbers and types of eclipsing binaries detectable in the idealised
SuperWASP survey and examine their dependence on the initial
mass ratio or secondary mass distribution. By considering the ratio of
the number of detectable eclipsing binaries to the number of
detectable single stars we also investigate if some regions in the
Galaxy are expected to produce more false planetary detections due to
eclipsing binaries than others. Our method of calculation is outlined
in some detail in Section 2, while the results are presented in
Sections 3 and 4. The final section is devoted to concluding remarks.

\section{Method}

\subsection{Eclipsing binaries detectable by SuperWASP}
\label{web}

As mentioned in the introduction, each camera on the SuperWASP fork
mount is capable of simultaneously monitoring several thousands of
7--13\,mag stars with a photometric precision better than
10\,mmag. In order to incorporate these instrument characteristics in
a binary population synthesis code, a transition needs to be made from
bolometric luminosities obtained from stellar evolution calculations
to apparent visual magnitudes seen by an observer. We make this
transition in the following steps.

First, we determine the absolute visual magnitudes $M_{{\rm V},1}$ and
$M_{{\rm V},2}$ of the binary components from their bolometric
luminosities $L_1$ and $L_2$ as
\begin{equation}
M_{{\rm V},i} = -2.5 \log {L_i \over L_\odot} - BC_i + M_{\rm bol, \odot},
  \,\,\, i=1,2, \label{amv}
\end{equation}
where $BC_i$ is the bolometric correction of component $i$, and
$M_{\rm bol, \odot}=4.76$\,mag is the bolometric magnitude of the
Sun. For stars with a hydrogen-rich envelope, the bolometric
correction is determined from the effective temperature by means of
polynomial fits to the empirical $BC$--$T_{\rm eff}$ relations given
by Flower (1996)\footnote{Corrections to the erroneous fits in Table~6
of Flower (1996) were kindly provided by Pasi Nurmi and Henri Boffin
(see also Nurmi \& Boffin 2003)}. For naked helium stars and white
dwarfs, the bolometric correction is assumed to be equal to -5.0
(Smith, Meynet \& Mermilliod 1994).

Secondly, we define the visual luminosity $L_{{\rm V},i}$ of
component $i$ as
\begin{equation}
\displaystyle {L_{{\rm V},i} \over L_\odot} = 10^{-\left( 
  M_{{\rm V},i} - M_{{\rm V},\odot} \right)/2.5},  \label{Lv}
\end{equation}
where $M_{{\rm V},\odot}$ is the absolute visual magnitude of the Sun.
For consistency, we determine $M_{{\rm V},\odot}$ from Eq.~(\ref{amv})
using the same $BC$--$T_{\rm eff}$ relations as used to determine
$M_{{\rm V},1}$ and $M_{{\rm V},2}$. It follows that $M_{{\rm
      V},\odot} = 4.84$\,mag\footnote{This is slightly larger than
  the usual $M_{{\rm V},\odot} = 4.83$\,mag due to the adopted $M_{\rm
    bol, \odot}$ and $BC$--$T_{\rm eff}$ relations which yield
$BC_\odot=-0.0778$. Since this affects the distance to which a star of
a given magnitude is visible by less than 0.5\%, the adopted $M_{{\rm
    V},\odot}$ does not significantly affect any of the results
presented in this investigation.}. The added luminosity of both 
binary components then gives rise to an absolute visual binary
magnitude
\begin{equation}
M_{\rm V,binary} = \displaystyle M_{{\rm V},\odot} - 2.5\, \log 
  {{L_{{\rm V},1} + L_{{\rm V},2}} \over L_\odot}.  \label{amvb}
\end{equation}

Thirdly, we determine the apparent visual magnitude of the binary as
\begin{equation}
m_{\rm V,binary} = M_{\rm V,binary} + 5\, \log_{10} 
  \left( d/{\rm pc} \right) - 5 + A_{\rm V}(d,l,b).  \label{mv}
\end{equation}
Here $d$ is the distance of the binary from the Sun, $l$ and
$b$ are the binary's heliocentric Galactic longitude and latitude, and
$A_{\rm V}(d,l,b)$ is the apparent increase in the visual magnitude of
the binary caused by interstellar extinction along the
line-of-sight. We incorporate the dependence of $A_{\rm V}$ on $d$,
$l$, and $b$ in the calculations by means of the extinction model of
Hakkila (1997). The model combines the results of several independent
extinction studies and also corrects some of the systematic biases
these studies are subjected to (see Hakkila 1997 for details and
references).

Fourthly, during an eclipse, the apparent visual magnitude of an
eclipsing binary increases by  
\begin{equation}
\Delta m_{\rm V, binary} = -2.5 \log 
  \left( 1 - {{\Delta L_{\rm V}} 
  \over {L_{{\rm V},1} + L_{{\rm V},2}}} \right),  \label{dm}
\end{equation}
where $\Delta L_{\rm V}$ is the decrease of the total visual
luminosity $L_{{\rm V},1} + L_{{\rm V},2}$. If the radius $R_1$ of the
primary is larger than the radius $R_2$ of the secondary, the decrease
in luminosity during a {\em total} eclipse is given by
\begin{equation}
\Delta L_{\rm V} = \max \left[ \left( {R_2 \over R_1} 
  \right)^2 L_{{\rm V}, 1}, L_{{\rm V}, 2} \right].  \label{Lec}
\end{equation}
The decrease in luminosity for $R_2>R_1$ is obtained by exchanging the
indices "1" and "2" in Eq.~(\ref{Lec}). Note that both $\Delta L_{\rm
  V}$ and $\Delta m_{\rm V, binary}$ are defined as positive
quantities.

\subsection{The Galactic disc model}

The stellar and binary content of the Galaxy are usually modelled by
superpositions of density laws describing different Galactic
components such as the bulge, the disc, and the halo (see, e.g.,
Bahcall 1986 for a review). For the purpose of this investigation, we
only consider the dominant component made up by the Galactic disc,
which we model by means of a stellar distribution law of the form
\begin{equation}
n(R,z) = n_0 \exp \left( - {R \over {h_{\rm R}}} \right) \exp 
  \left( - {{|z|} \over {h_{\rm z}}} \right).  \label{disc}
\end{equation}
Here, $n_0$ is a normalisation constant, $R$ and $z$ are the disc's
natural cylindrical coordinates, $h_{\rm R}$ is the disc's scale
length, and $h_{\rm z}$ is its scale height. We normalise the
distribution to unity by setting $n_0 = 1/\!\left( 4 \pi h_{\rm z}
h_{\rm R}^2 \right)$.

Gilmore \& Reid (1983) pointed out that star counts towards the
Southern Galactic pole were not well fitted by a single vertical
exponential, but that a good agreement could be obtained when two such
exponentials, representing the so-called thin and thick disc, with
different scale heights were considered. The thin disc has a scale
height of about 300\,pc and dominates up to $z \approx 1\, {\rm
kpc}$. Its stars cover a wide range of stellar ages between 0 and
10\,Gyr. The properties of the thick disc, on the other hand, are less
well constrained. They are often referred to as "intermediate" between
those of the Galactic disc and the Galactic halo. In particular, the
abundances of thick disc stars point towards a moderately metal-poor
stellar population. A more extreme metal-weak thick disc, possibly
being the tail of the thick disc, also exists, but its contribution to
the structure and the content of the Galactic disc is much smaller
than that of the thin and thick disc. Hence, as a first approximation,
we here only consider binaries with solar metallicity component
stars. Estimates for the scale height of the thick disc range from 500
to 1500\,pc (e.g. Norris 1999, Du et al. 2003; and references
therein). Much larger scale heights have also been claimed, but seem
to be ruled out by recent scale height determinations from HST
photometry (Kerber, Javiel \& Santiago 2001) and 2MASS data (Ojha
2001, Cabrera-Lavers, Garz{\' o}n, \& Hammersley 2005). The age of the
thick disc is furthermore thought to be in the range of 12-15\,Gyr, so
that it is relatively old in comparison to the thin disc.

Observations supporting a large spread in the thick disc's age
comparable to that of the thin disc, or the existence of an age gap
between the thin and thick disc have so far been inconclusive (Liu \&
Chaboyer 2000). We therefore assume that the Galactic disc consists of
a thin and thick component which originate from different
non-overlapping star formation epochs and that the thin disc formed
immediately after the thick disc. The total age of the disc is
assumed to be 13\,Gyr. Binaries formed during the last 10\,Gyr of
star formation in the Galactic disc are distributed according to a
double exponential with $h_{\rm R}=2.8\, {\rm kpc}$ and $h_{\rm
z}=300\, {\rm pc}$, while binaries formed between 10\,Gyr and 13\,Gyr
ago are distributed according to a double exponential with $h_{\rm
R}=3.7\, {\rm kpc}$ and $h_{\rm z}=1\, {\rm kpc}$. This gives rise to
\begin{enumerate}
\item a population of stars with ages between 0 and 10\,Gyr and a
relatively small vertical scale height, representative of the thin
disc stars, and
\item a population of stars with ages of 10--13\,Gyr and a much larger
vertical scale height, representative of the thick disc stars.
\end{enumerate}
Hence, over a time span of 13\,Gyr, we form 10/3 times more stars in
the thin disc than in the thick disc. As we show, below these star
formation histories and the adopted thin and thick distributions given
by Eq.~(\ref{disc}) result in stellar densities in the thin and thick
Galactic disc that are compatible with current observational estimates
(e.g. Norris 2001).  For more details on the adopted scale lengths and
scale heights, we refer to Kerber et al. (2001), Ojha (2001), Siegel
et al. (2003), and Du et al. (2003).

\subsection{Binary population synthesis}

We use the BiSEPS binary population synthesis code described by
Willems \& Kolb (2002, 2004) to construct present-day populations of
eclipsing binaries in the Galactic disc detectable by the idealised
SuperWASP survey.  We
recall that single star evolution is treated using the formulae
derived by Hurley, Pols \& Tout (2000), binary orbits are assumed to
be circular, and rotational angular velocities are kept synchronised
with the orbital motion at all times. Orbital angular momentum losses
via gravitational radiation and/or magnetic braking are taken into
account as in Hurley, Tout \& Pols (2002), and Roche-lobe overflow is
treated as in Willems \& Kolb (2004). All calculations presented in
this paper are furthermore based on the binary evolution parameters
corresponding to the standard model (model~A) of Willems \& Kolb
(2004).

The evolution of a large number of binaries consisting of two zero-age
main-sequence components was followed up to a maximum evolutionary age
of 13\,Gyr. The initial primary and secondary masses, $M_1$ and $M_2$,
are taken from a grid of 50 logarithmically spaced masses in the
interval between $0.1$ and $20\,M_\odot$, and the initial orbital
periods, $P_{\rm orb}$, from a grid of 250 logarithmically spaced
periods in the interval between $0.1$ and $10\,000$ days.
In order to avoid spurious spikes in the distribution functions of
simulated binaries, each grid point is actually randomly offset from
the center of the grid cells according to uniform $\log M_1$, $\log M_2$, 
and $\log P_{\rm orb}$ distributions within the boundaries of the grid 
cells (cf. Hurley et al. 2002). 
For symmetry reasons only binaries with $M_1 > M_2$ are 
evolved. At the end of the calculation of each binary sequence, the
evolutionary track is scanned for phases where {\it total} eclipses
would be detectable by SuperWASP. We furthermore only retain binaries
that are detached or in a thermally and dynamically stable
mass-transfer phase.  In view of the generally short life time of
thermal and dynamical time scale mass-transfer episodes, we do not
expect this limitation to significantly affect our results.

The statistical contribution of a binary to the population of
eclipsing binaries detectable by SuperWASP is determined by the
probability distribution functions of the initial binary parameters,
the adopted star formation rate, the fraction of stars in binaries,
the position of the binary in the Galactic disc, and the probability
that the orbit is oriented suitably for the observation of total
eclipses.  We assume the initial primary masses to be distributed
according to an initial mass function (IMF) with a slope of -1.3 for
$0.1\, M_\odot \le M_1 < 0.5\, M_\odot$, -2.2 for $0.5\, M_\odot \le
M_1 < 1.0\, M_\odot$, and -2.7 for $1.0\, M_\odot \le M_1$. We
furthermore assume that the distribution of initial orbital
separations is logarithmically flat, such that $\chi (\log a) =
0.078636$ for $3 \le a/R_\odot\le 10^6$. The initial mass ratio
distribution is parametrized as $n(q) \propto q^s$ with $s \in \bbbr$
a free parameter and $0 < q=M_2/M_1 \le 1$, though we also investigate
the case where both binary components are picked independently from
the same IMF (see Willems \& Kolb 2002 for details and references).
We furthermore assume that 50\% of all stars are binaries and that the
Galactic star-formation rate has been constant for the past
13\,Gyr. The star-formation is calibrated so that one binary or single
star with $M_1 > 0.8\, M_\odot$ is born in the Galaxy each year, in
concordance with the observationally inferred birthrate of Galactic
white dwarfs (Weidemann 1990).

The role of the position of the binary in the Galactic disc is taken
into account by means of Eqs. (\ref{mv}) and (\ref{disc}). More
specifically, for a given absolute visual magnitude $M_{\rm V,binary}$
and a given set of Galactic coordinates $(l,b)$, Eq.~(\ref{mv})
translates the optimal SuperWASP magnitude range $7\,{\rm mag} \le
m_{\rm V,binary} \le 13\,{\rm mag}$ into a range of optimal distances
$d_{\rm min} \le d \le d_{\rm max}$. Since the translation depends on
the interstellar extinction coefficient $A_V(d,l,b)$, which itself is
a function of the distance $d$ of the binary from the Sun, the optimal
distance range has to be determined iteratively. The dependence of
$A_{\rm V}$ on the Galactic longitude and latitude furthermore implies
that $d_{\rm min}$ and $d_{\rm max}$ are also functions of $l$ and
$b$. The statistical weight of a binary with Galactic coordinates
$(l_0,b_0)$ is then determined by (i) integrating Eq.~(\ref{disc})
over all distances $d$ ranging from $d_{\rm min}$ to $d_{\rm max}$,
and (ii) integrating the resulting two-dimensional function of $l$ and
$b$ over a rectangular field of view centred on $(l_0,b_0)$. To this
end, the Galactocentric cylindrical coordinates $(R,z)$ appearing in
Eq.~(\ref{disc}) are expressed in terms of the Heliocentric spherical
coordinates $(d,l,b)$ as
\begin{equation}
\renewcommand{\arraystretch}{1.5}
\left.
\begin{array}{l c l}
R & = & \left( d^2 \cos^2 b - 2 d R_\odot \cos b \cos l
  + R_\odot^2 \right)^{1/2}, \\
z & = & d \sin b + z_\odot.
\end{array}
\right\} \label{trans}
\end{equation}
Here $R_\odot$ is the distance of the projection of the Sun on the
Galactic plane to the Galactic centre, and $z_\odot$ is the height of
the Sun above the Galactic plane. We here adopt the values $R_\odot =
8.5\,{\rm kpc}$ and $z_\odot = 30\,{\rm pc}$ (e.g. Binney \&
  Tremaine 1987; Reid 1993; Chen et al. 2001; and references therein).

The probability to observe binary eclipses, finally, is determined by
the radii of the component stars and the orbital separation. In
particular, two binary components with radii $R_1$ and $R_2$ separated
by a distance $a$, will show total eclipses for orbital inclinations
$i \ge i_0$, with $i_0$ determined by
\begin{equation}
\cos i_0 = {{|R_1-R_2|} \over a}.  \label{proj}
\end{equation}
Adopting a uniform distribution for the cosine of the orbital
inclination $i$ then gives an eclipse probability given by 
\begin{equation}
P(i \ge i_0) = \int_{i_0}^{\pi/2} \sin i\, di = 
  {{|R_1-R_2|} \over a}.  \label{ecl}
\end{equation}

\section{Galactic distributions}

\begin{table}
\caption{Idealised transit survey characteristics.}
\label{models}
\begin{tabular}{ccc}
\hline
 model & magnitude range & photometric precision (mag) \\
\hline
1 & 7--13 & 0.01 \\
2 & 7--15 & 0.01 \\
3 & 7--13 &  0.1 \\
4 & 7--15 &  0.1 \\
\hline
\end{tabular}
\end{table}

We used the technique and the assumptions outlined in the previous
section to construct Galactic distributions of eclipsing binaries and
single stars detectable by an idealised SuperWASP survey.
Specifically, we adopt four different idealised transit survey
characteristics. In survey model~1 (our standard survey model), the
magnitude range and eclipse detection threshold are assumed to be
7--13\,mag and 0.01\,mag, respectively. The assumptions adopted in the
other three survey models are summarised in Table~\ref{models}.  We
note that the eclipse detection threshold is irrelevant for single
stars, so that models 3 and 4 are equivalent to models 1 and 2 for
these stars.

\begin{figure}
\resizebox{\hsize}{!}{\includegraphics{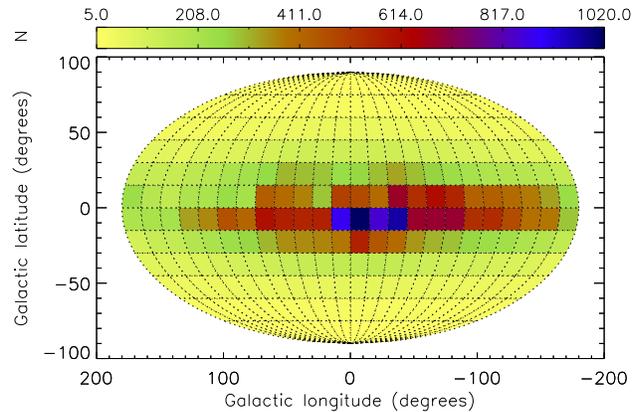}} \\
\caption{Distribution of the number $N$ of eclipsing binaries
detectable in the Galactic disc, for survey model~1 and a flat initial
mass ratio distribution $n(q)=1$. Note that the color scale for $N$ is
linear.}
\label{WEB2}
\end{figure}

The Galactic distribution of eclipsing binaries detectable by
SuperWASP is shown in Fig.~\ref{WEB2}, for survey model~1 and a flat
initial mass ratio distribution $n(q)=1$. The systems are grouped in
$15^\circ \times 15^\circ$ bins in Galactic longitude and latitude,
corresponding approximately to the combined field of view of four
SuperWASP cameras operating simultaneously in a rectangularly mounted
configuration. The double exponential distribution functions adopted
for the thin and thick Galactic disc components yield a strong
concentration of systems towards the Galactic plane ($b=0^\circ$) and
towards the Galactic centre ($l=0^\circ$, $b=0^\circ$). The asymmetry
in the distributions with respect to the Galactic centre is caused by
the Galactic longitude and latitude dependency of the interstellar
extinction model [see, e.g., Fig.~11 in Hakkila et al. (1997)]. The
largest number of systems is found in the $15^\circ \times 15^\circ$
field of view centred on $(l,b) = (-7.5^\circ, -7.5^\circ)$. The
behaviour of the distributions of detectable eclipsing binaries for
the other three survey models is similar to that of model~1. The same
applies to the Galactic distributions of single stars detectable by
SuperWASP.

\begin{figure}
\resizebox{\hsize}{!}{\includegraphics{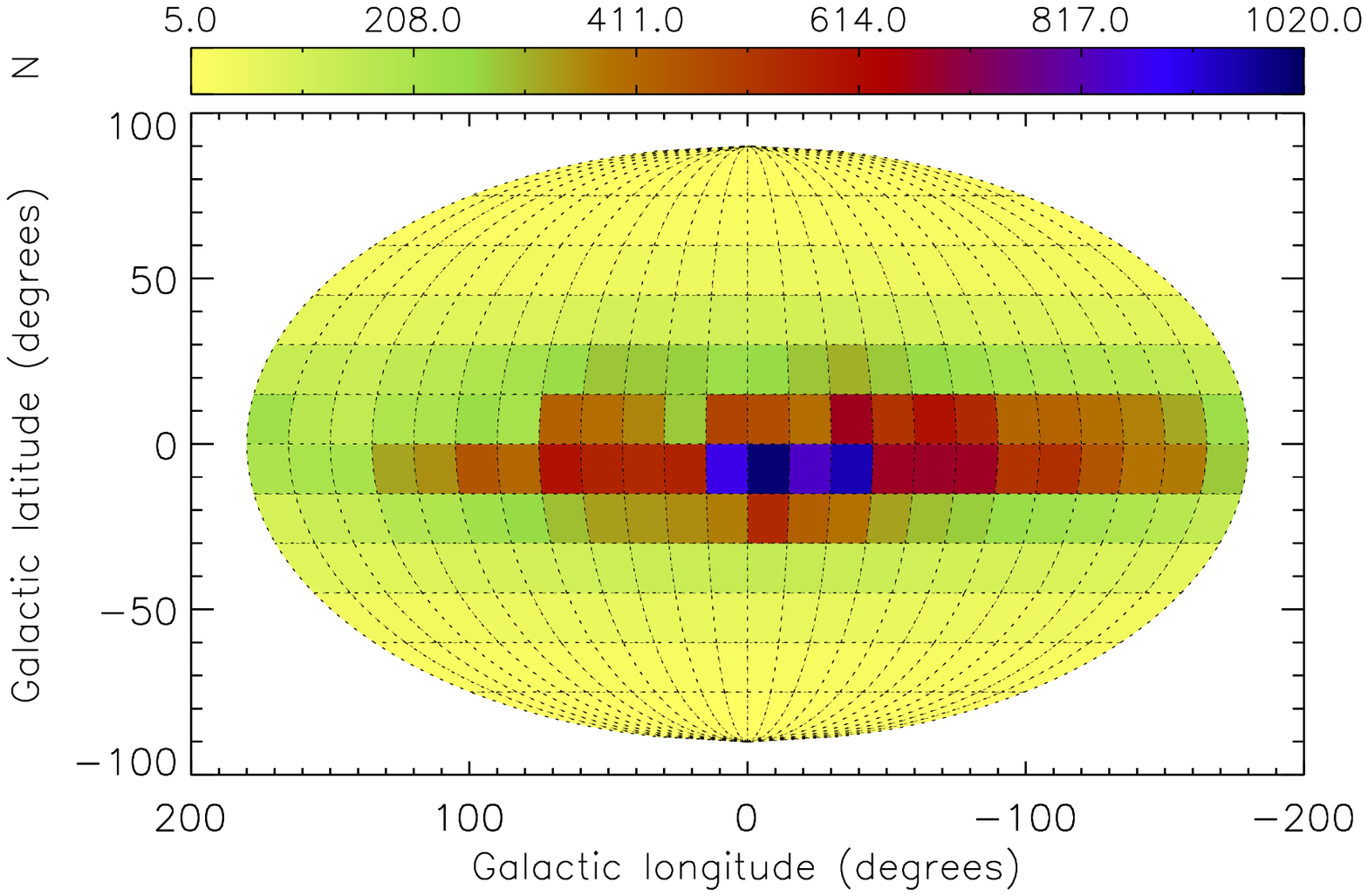}} \\
\resizebox{\hsize}{!}{\includegraphics{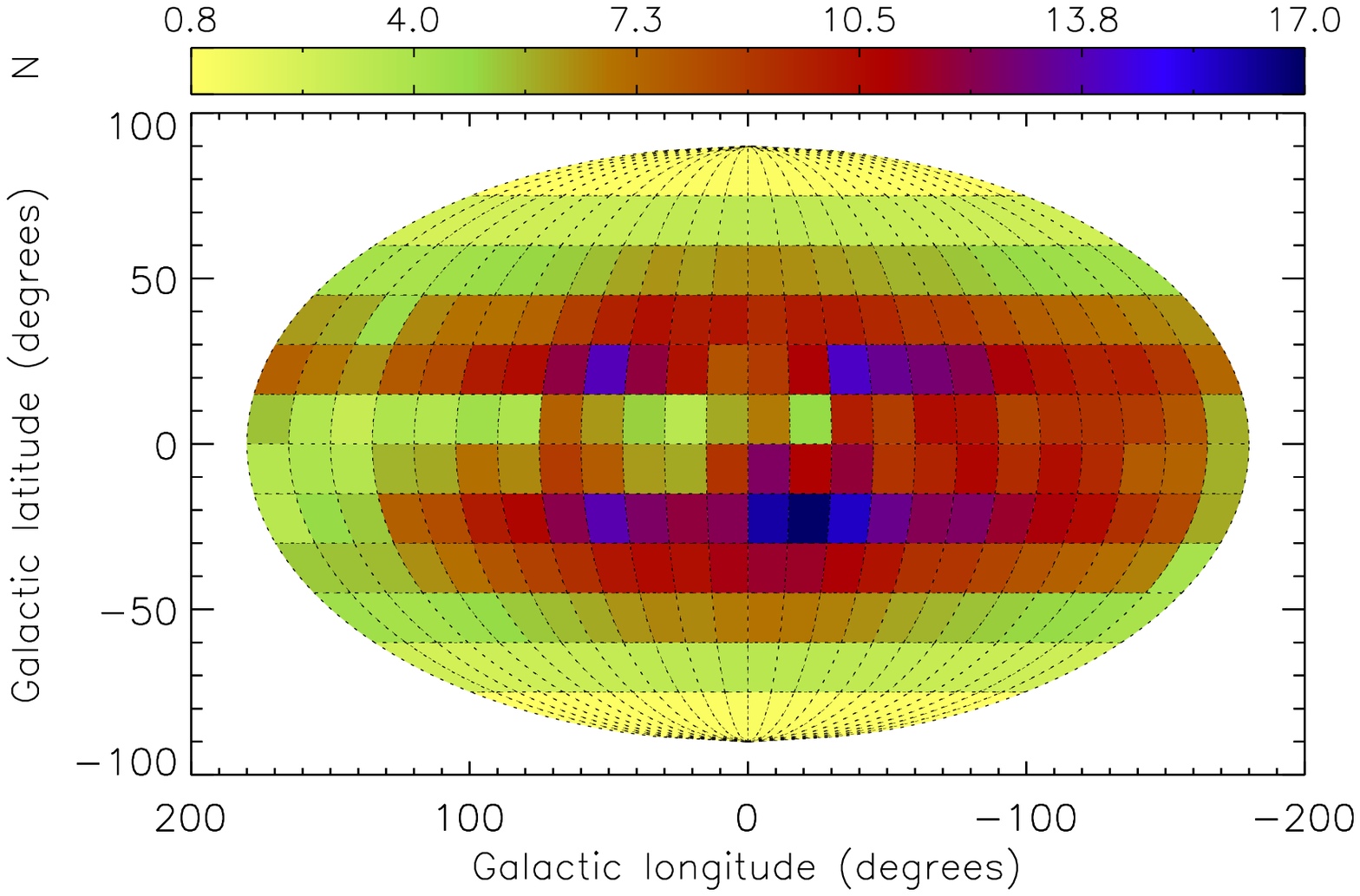}} \\
\caption{As Fig.~\ref{WEB2}, but with the Galactic disc separated into
  its thin (top panel) and thick (bottom panel) components.}
\label{WEB1}
\end{figure}

The contributions of the thin and thick Galactic disc to the
population of eclipsing binaries detectable by SuperWASP are shown
separately in Fig.~\ref{WEB1}. The thin disc component contains
approximately 96\% of the sample of detectable eclipsing binaries
(see Section~\ref{snum}). The fine structure in the thick
disc component reflects the variations of the interstellar extinction
as a function of Galactic longitude and latitude. These variations are
more apparent for the thick Galactic disc than for the thin Galactic
disc because it is much less centrally condensed.

\begin{figure}
\resizebox{\hsize}{!}{\includegraphics{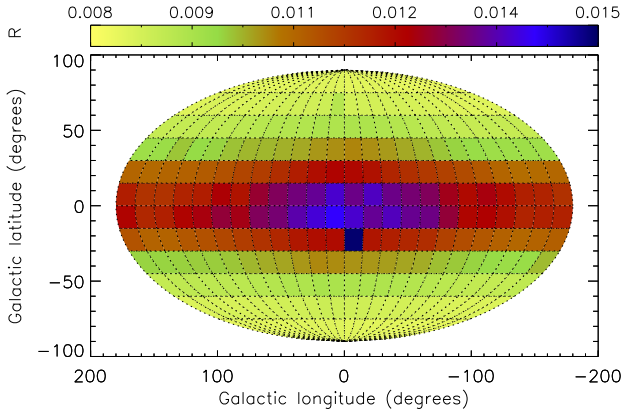}}
\caption{As Fig.~\ref{WEB2}, but in units of the number of detectable
  single stars.}
\label{ratio}
\end{figure}

In Fig.~\ref{ratio}, we show the variation of the ratio $R$ of the
number of eclipsing binaries to the number of single stars detectable
by SuperWASP as a function of Galactic longitude and latitude, in the
case of survey model~1 and a flat initial mass ratio distribution. The
ratio is largest for the bin centred on $(l,b) = (-7.5^\circ,
-22.5^\circ)$, so that this region is less favourable for SuperWASP if
false exoplanet transit detections due to eclipsing binaries are to be
minimized. This fairly localized peak in the distribution of $R$ is 
already apparent as a local maximum in $l$ direction in Fig.~\ref{WEB2}.
The peak results from the interplay between a decrease in the interstellar
extinction in this bib (see Fig.~11 in Hakkila 1997) and the non-trivial 
differential behaviour between single and binary stars to the interstellar 
extinction (binaries emit the combined light of two stars, 
affecting the distance to which they are observable).
For $75^\circ \la l \la 285^\circ$,
more favourable regions are found at any Galactic latitude between
$0^\circ$ and $90^\circ$, while for $l \la 75^\circ$ and $l \ga
285^\circ$ more favourable regions are located at Galactic latitudes
$|b| \ga 15^\circ$ (except for the bin centred on $l=-7.5^\circ$ and
$b=-22.5^\circ$). The same holds true for survey models 2--4. In all
models, the ratio $R$ varies across the sky by less than a factor of
2.5.

\section{Absolute and relative numbers}
\label{snum}

In the previous Section, we have seen that the distribution of
eclipsing binaries and single stars detectable by an idealised
SuperWASP survey in the Galactic disc is similar for all four survey
models listed in Table~\ref{models}. In this section we show the
effect of the model assumptions on the number of detectable eclipsing
binaries and single stars, and on their breakdown in subgroups
according to the type of stars involved.

The total number of detectable eclipsing binaries and single stars is
listed in Table~\ref{num1}, for each of the four survey models
considered and for different initial mass ratio or initial secondary
mass distributions. For survey model~1 and a flat initial mass ratio
distribution $n(q)=1$, the number of eclipsing binaries is of the
order of $4 \times 10^4$. Extending the detectable magnitude range
from 7--13\,mag (model~1) to 7--15\,mag (model~2) increases the total
number of systems by approximately a factor of 4. Increasing the
minimum detectable eclipse depth from 0.01\,mag (model~1) to 0.1\,mag
(model~3), on the other hand, yields a decrease of the total number of
systems by about a factor of 2. The initial mass ratio or
secondary mass distribution has a significant effect on the absolute
number of systems only when small initial mass ratios $M_2/M_1$ are
favoured: when $n(q) \propto q^{-0.99}$ the number of detectable
eclipsing binaries decreases by approximately a factor of 50
compared to the case of $n(q)=1$.  The dominance of extreme-mass ratio
systems implies a large number of undetectable small-amplitude eclipses.

\begin{table}
\caption{The total number of eclipsing binaries and single stars
  detectable by an idealised SuperWASP survey, for each of the four
  survey models listed 
  in Table~\ref{models} and for different initial mass ratio or
  initial secondary mass distributions. The last column in the table
  lists the ratio of the total number of detectable eclipsing binaries
  to the total number of detectable single stars. }
\label{num1}
\begin{tabular}{cccc}
\hline
 model & eclipsing binaries & single stars & ratio \\
\hline
\hline
\multicolumn{4}{c}{$n(q) \propto q$, $0 < q \le 1$} \\
\hline
1 & $4.8 \times 10^4$ & $3.7 \times 10^6$ & $1.3 \times 10^{-2}$ \\
2 & $2.0 \times 10^5$ & $1.8 \times 10^7$ & $1.1 \times 10^{-2}$ \\
3 & $2.4 \times 10^4$ & $3.7 \times 10^6$ & $6.5 \times 10^{-3}$ \\
4 & $1.0 \times 10^5$ & $1.8 \times 10^7$ & $5.6 \times 10^{-3}$ \\
\hline
\hline
\multicolumn{4}{c}{$n(q) = 1$, $0 < q \le 1$} \\
\hline
1 & $4.3 \times 10^4$ & $3.7 \times 10^6$ & $1.2 \times 10^{-2}$ \\
2 & $1.8 \times 10^5$ & $1.8 \times 10^7$ & $1.0 \times 10^{-2}$ \\
3 & $1.8 \times 10^4$ & $3.7 \times 10^6$ & $4.9 \times 10^{-3}$ \\
4 & $7.9 \times 10^4$ & $1.8 \times 10^7$ & $4.4 \times 10^{-3}$ \\
\hline
\hline
\multicolumn{4}{c}{$n(q) \propto q^{-0.99}$, $0 < q \le 1$} \\
\hline
1 & $1.1 \times 10^3$ & $3.7 \times 10^6$ & $3.0 \times 10^{-4}$ \\
2 & $4.9 \times 10^3$ & $1.8 \times 10^7$ & $2.7 \times 10^{-4}$ \\
3 & $3.0 \times 10^2$ & $3.7 \times 10^6$ & $8.1 \times 10^{-5}$ \\
4 & $1.4 \times 10^3$ & $1.8 \times 10^7$ & $7.8 \times 10^{-5}$ \\
\hline
\hline
\multicolumn{4}{c}{$M_2$ from same IMF as $M_1$} \\
\hline
1 & $3.2 \times 10^4$ & $3.7 \times 10^6$ & $8.6 \times 10^{-3}$ \\
2 & $1.5 \times 10^5$ & $1.8 \times 10^7$ & $8.3 \times 10^{-3}$ \\
3 & $4.6 \times 10^3$ & $3.7 \times 10^6$ & $1.2 \times 10^{-3}$ \\
4 & $2.5 \times 10^4$ & $1.8 \times 10^7$ & $1.4 \times 10^{-3}$ \\
\hline
\end{tabular}
\end{table}

The total number of detectable single stars is typically more than two
orders of magnitude larger than the total number of detectable
eclipsing binaries in the same survey model (the ratio of the number
of detectable eclipsing binaries to the number of detectable single
stars is listed in the last column of Table~\ref{num1}). In the case
of the initial mass ratio distribution $n(q) \propto q^{-0.99}$, the
difference even amounts to three to four orders of magnitude,
depending on the adopted survey model. The dominance of single stars
is partly because eclipsing binaries are subjected to the additional
observational threshold that their eclipses must be deep enough to be
observable and identifiable by the transit survey, and partly because
the presence of a companion can alter the life time and evolutionary
path of a binary component in comparison to those of a single
star. Mass transfer, e.g., can strip a red giant down to its
  core, leaving behind a rapidly dimming white dwarf which is less likely
  to be detected than its giant progenitor.
We also note that the number of detectable eclipsing binaries given in
Table~\ref{num1} is by no means representative of the total number of
binaries seen by the survey since the latter includes eclipsing
systems with undetectable eclipse amplitudes as well as non-eclipsing
binaries.

In Table~\ref{num3}, the population of detectable single stars is 
subdivided according to the type of star and according
to their membership to the thin or thick Galactic disc. Approximately
60\% of all single stars in the
magnitude range between 7 and 13\,mag (model~1) are main-sequence
stars, 20\% are horizontal-branch stars, and 10\% are
giant-branch stars, 2\% are Hertzsprung-gap stars, 1\%
are asymptotic giant branch stars, and 1\% are white dwarf
stars. If the detectable magnitude range is extended from 7-13\,mag to
7-15\,mag (model~2), the relative number of horizontal branch stars
decreases somewhat, while the relative numbers of main-sequence and
white dwarf stars increases. The ratio of the number of detectable
single stars in the thin Galactic disc to the number of detectable
single stars in the thick Galactic disc is approximately 15 to 1 for
survey model~1 and 10 to 1 for survey model~2, comparable to current
estimates for the ratio of stellar densities in the thin and thick
Galactic disc (e.g. Norris 2001). 

\begin{table}
\caption{Contributions of different types of stars in the thin and
thick Galactic disc to the population of detectable single stars,
for survey models~1 and~2. In column 1, MS
stands for main sequence, HG for Hertzsprung gap, GB for giant branch,
HB for horizontal branch, AGB for asymptotic giant branch, and WD for
white dwarf.}
\label{num3}
\begin{tabular}{lcccc}
\hline
  & \multicolumn{2}{c}{model 1} & \multicolumn{2}{c}{model 2} \\
 stellar type & thin disc & thick disc & thin disc & thick disc \\
\hline
MS         & 60.76\% & 0.78\% & 64.83\% & 1.91\% \\
HG         &  1.70\% & 0.17\% &  1.85\% & 0.36\% \\
GB         &  9.84\% & 2.61\% &  8.17\% & 3.77\% \\
HB         & 20.02\% & 1.92\% & 13.87\% & 1.89\% \\
AGB        &  0.96\% & 0.10\% &  0.66\% & 0.09\% \\
WD         &  1.00\% & 0.14\% &  2.22\% & 0.38\% \\
Total      & 94.28\% & 5.72\% & 91.60\% & 8.40\% \\
\hline
\end{tabular}
\end{table}

\begin{figure*}
\resizebox{8.0cm}{!}{\includegraphics{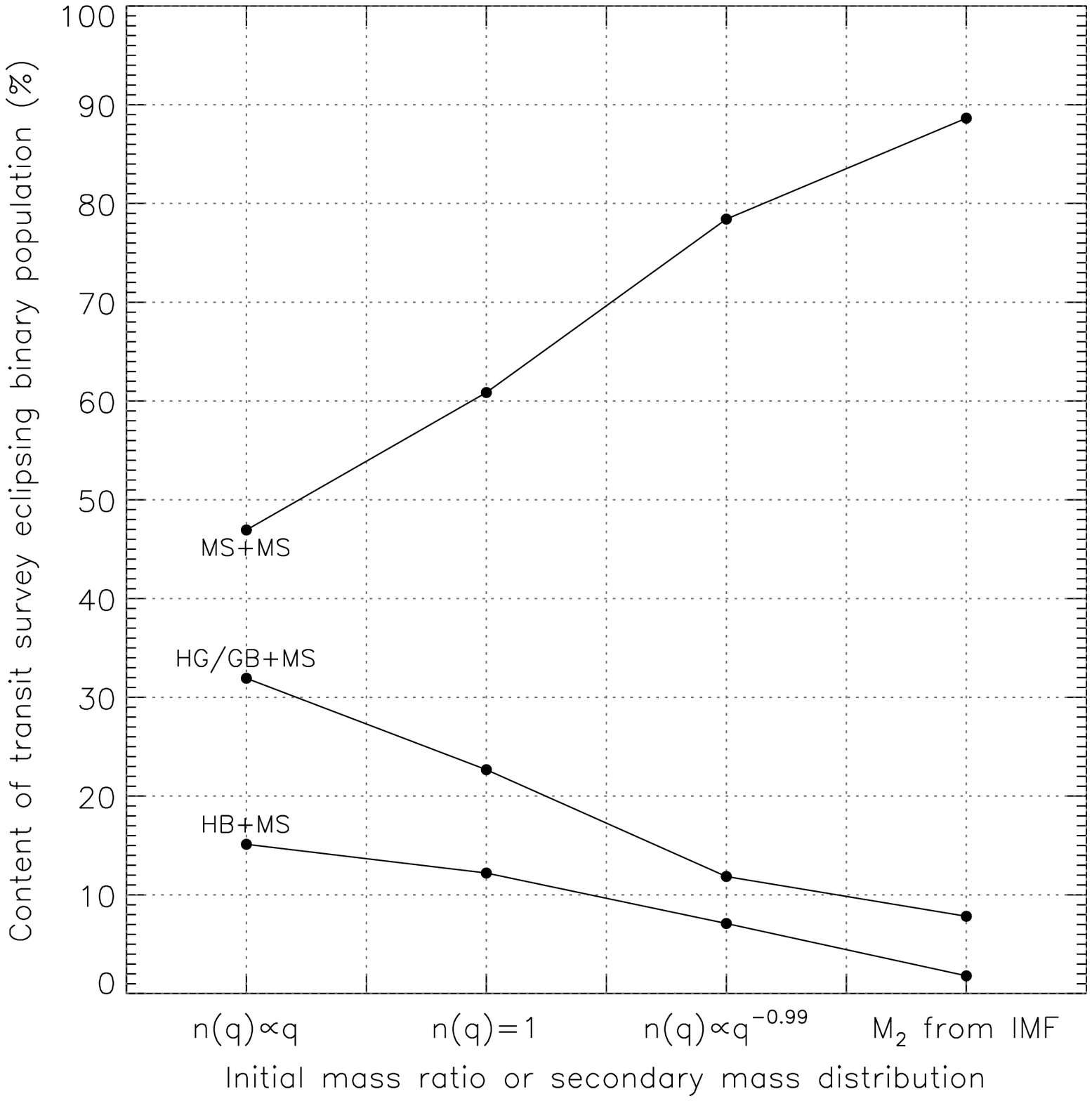}}
\resizebox{8.0cm}{!}{\includegraphics{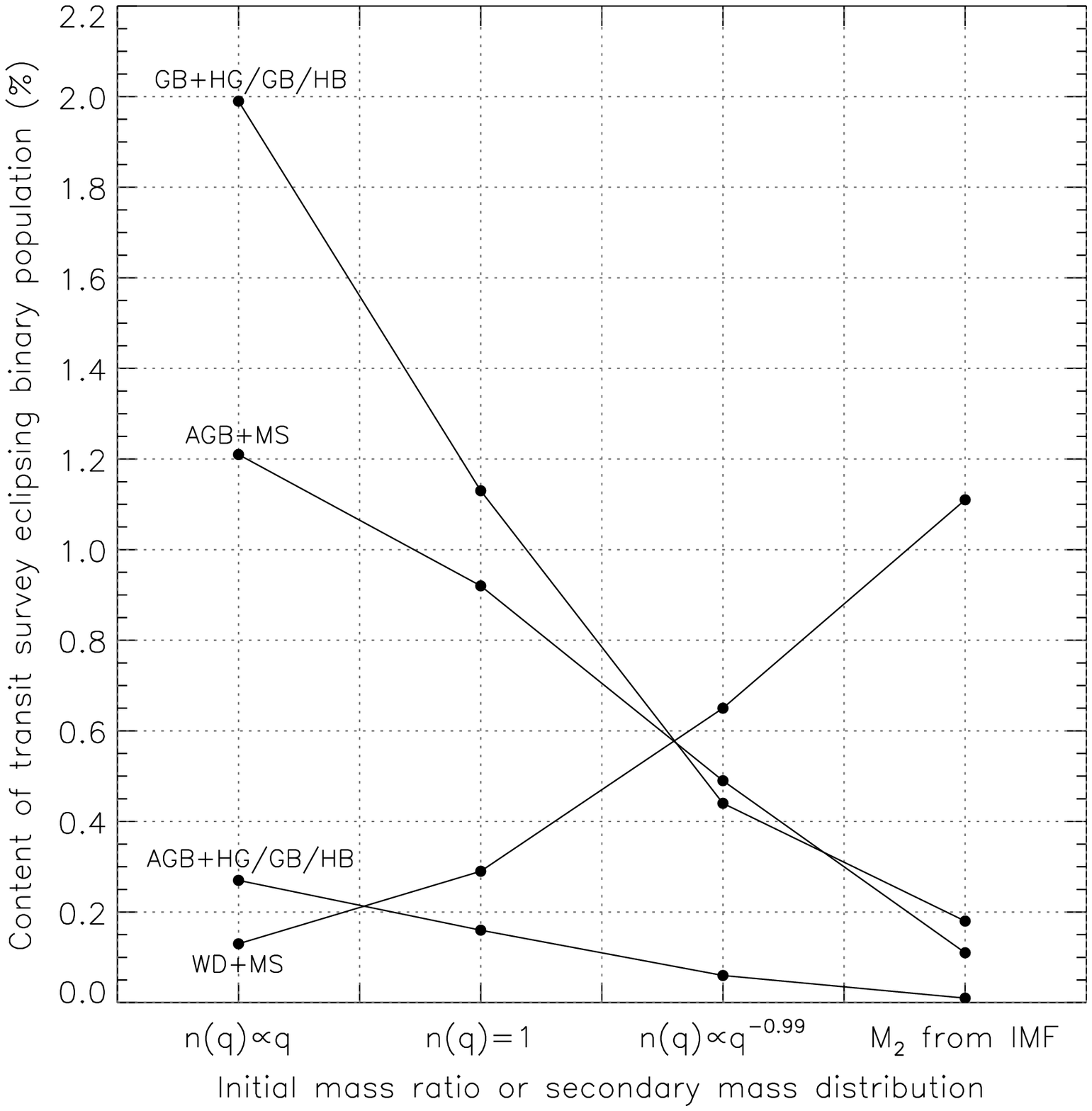}} \\
\caption{Relative contribution of the most common types of binaries to
the population of eclipsing binaries detectable by an idealised
SuperWASP survey in the
(thin and thick) Galactic disc, for survey model~1 and different
initial mass ratio or initial secondary mass distributions. The
abbreviations MS, HG, GB, ... have the same meaning as in
Tables~\ref{num3} and~\ref{num2}.} 
\label{pop}
\end{figure*}

Table~\ref{num2} gives a detailed breakdown into types of binaries
that make up the sample of eclipsing binaries detectable by the
transit survey. In the case of survey model~1, the dependence of the
relative contributions of different types of systems on the adopted
initial mass ratio or initial secondary mass distribution is
visualized in Fig.~\ref{pop}.\footnote{Note that Kolb \& Willems
(2004) presented early results of these simulations. The fractions in
their Table 2 are now superseded by the results presented here.}

The dominant group of detectable eclipsing binaries consists of two
detached main-sequence stars. In the case of survey model~1 and a flat
initial mass ratio distribution, this subgroup of systems accounts for
60\% of the population of detectable eclipsing binaries. Detached
giant-branch main-sequence star binaries account for approximately
20\% of the population, detached horizontal-branch main-sequence star
binaries for approximately 10\%, and Hertzsprung-gap main-sequence
star binaries for about 2\%. Other types of binaries contribute at
most 1\%. The relative number of detached double main-sequence star
binaries increases to 65\% for survey model~2 and 70--80\% for survey
models~3 and~4. The contribution of giant-branch main-sequence star
binaries, on the other hand, decreases from 20\% in survey models~1
and~2 to 5\% in survey models~3 and~4.  The relative number of
detectable eclipsing binaries in the thin Galactic disc furthermore
varies from 93\% and 99\% between the different survey models
considered. In addition, it is interesting to note that the most
abundant group of eclipsing binaries detectable by the transit survey
model in the thick Galactic disc consists of detached giant-branch
main-sequence star binaries.  This is because essentially all former
double main-sequence stars with large-amplitude eclipses have evolved
off the main sequence in the thick disc population.

\begin{table*}
\caption{Contributions of the most common types of binaries in the
thin and thick Galactic disc to the population of eclipsing binaries
detectable by an idealised SuperWASP survey, for each of the survey
models listed in Table~\ref{models} and different initial mass ratio
or initial secondary mass distributions. In column 1, MS stands for
main sequence, HG for Hertzsprung gap, GB for giant branch, HB for
horizontal branch, AGB for asymptotic giant branch, nHe for naked
helium star, and WD for white dwarf. Detached and semi-detached
systems are indicated in column 2 by the labels D and SD,
respectively.}
\label{num2}
\begin{tabular}{lccccccccc}
\hline
binary & binary & \multicolumn{2}{c}{model 1} & \multicolumn{2}{c}{model 2} & \multicolumn{2}{c}{model 3} & \multicolumn{2}{c}{model 4} \\
components & type & thin disc & thick disc & thin disc & thick disc & thin disc & thick disc & thin disc & thick disc \\
\hline
\hline
\multicolumn{10}{c}{$n(q) \propto q$, $0 < q \le 1$} \\
\hline
MS+MS & D        &    46.72\% &    0.22\% &    50.05\% &    0.62\% &    67.30\% &    0.35\% &    70.93\% &    0.94\% \\
HG+MS & D        &     1.53\% &    0.11\% &     1.76\% &    0.28\% &     1.75\% &    0.14\% &     1.92\% &    0.33\% \\
GB+MS & D        &    25.30\% &    4.99\% &    22.88\% &    7.66\% &     7.40\% &    0.61\% &     7.02\% &    1.18\% \\
GB+HG/GB/HB & D  &     1.72\% &    0.27\% &     1.46\% &    0.35\% &     1.80\% &    0.42\% &     1.52\% &    0.52\% \\
HB+MS & D        &    15.06\% &    0.06\% &    11.56\% &    0.07\% &    15.04\% &    0.00\% &    11.51\% &    0.00\% \\
AGB+MS & D       &     1.20\% &    0.01\% &     0.92\% &    0.01\% &     0.42\% & $<$0.01\% &     0.31\% & $<$0.01\% \\
AGB+HG/GB/HB & D &     0.26\% &    0.01\% &     0.20\% &    0.01\% &     0.41\% & $<$0.01\% &     0.31\% & $<$0.01\% \\
nHe+MS & D       &     0.18\% &    0.00\% &     0.14\% &    0.00\% &     0.04\% &    0.00\% &     0.03\% &    0.00\% \\
WD+MS & D        &     0.13\% &    0.02\% &     0.25\% &    0.05\% &     0.08\% &    0.01\% &     0.15\% &    0.03\% \\
MS+MS & SD       &     0.65\% & $<$0.01\% &     0.53\% & $<$0.01\% &     1.30\% & $<$0.01\% &     1.02\% & $<$0.01\% \\
HG+MS & SD       &     0.12\% & $<$0.01\% &     0.10\% & $<$0.01\% &     0.23\% & $<$0.01\% &     0.19\% & $<$0.01\% \\
GB+MS & SD       &     0.91\% &    0.01\% &     0.73\% &    0.01\% &     1.81\% &    0.02\% &     1.41\% &    0.03\% \\
HB+MS & SD       &     0.11\% &    0.00\% &     0.06\% &    0.00\% &     0.15\% &    0.00\% &     0.09\% &    0.00\% \\
Other  &         &     0.40\% &    0.01\% &     0.30\% & $<$0.01\% &     0.73\% & $<$0.01\% &     0.56\% & $<$0.01\% 
\medskip \\
Total  &         &    94.29\% &    5.71\% &    90.94\% &    9.06\% &    98.46\% &    1.54\% &    96.97\% &    3.03\% \\
\hline
\hline
\multicolumn{10}{c}{$n(q) = 1$, $0 < q \le 1$} \\
\hline
MS+MS & D        &    60.58\% &    0.27\% &    63.33\% &    0.74\% &    73.55\% &    0.38\% &    76.55\% &    1.01\% \\
HG+MS & D        &     1.56\% &    0.12\% &     1.79\% &    0.28\% &     1.49\% &    0.12\% &     1.61\% &    0.27\% \\
GB+MS & D        &    17.71\% &    3.28\% &    15.85\% &    5.03\% &     5.64\% &    0.44\% &     5.25\% &    0.84\% \\
GB+HG/GB/HB & D  &     0.98\% &    0.15\% &     0.81\% &    0.19\% &     1.22\% &    0.28\% &     1.01\% &    0.34\% \\
HB+MS & D        &    12.17\% &    0.04\% &     9.12\% &    0.04\% &    12.24\% &    0.00\% &     9.19\% &    0.00\% \\
AGB+MS & D       &     0.91\% &    0.01\% &     0.68\% &    0.01\% &     0.33\% & $<$0.01\% &     0.24\% & $<$0.01\% \\
AGB+HG/GB/HB & D &     0.15\% &    0.01\% &     0.11\% & $<$0.01\% &     0.29\% & $<$0.01\% &     0.21\% & $<$0.01\% \\
nHe+MS & D       &     0.14\% &    0.00\% &     0.10\% &    0.00\% &     0.05\% &    0.00\% &     0.04\% &    0.00\% \\
WD+MS & D        &     0.29\% &    0.04\% &     0.57\% &    0.11\% &     0.27\% &    0.04\% &     0.53\% &    0.10\% \\
MS+MS & SD       &     0.52\% & $<$0.01\% &     0.41\% & $<$0.01\% &     1.24\% & $<$0.01\% &     0.95\% & $<$0.01\% \\
HG+MS & SD       &     0.09\% & $<$0.01\% &     0.08\% & $<$0.01\% &     0.22\% & $<$0.01\% &     0.17\% & $<$0.01\% \\
GB+MS & SD       &     0.66\% &    0.01\% &     0.52\% &    0.01\% &     1.58\% &    0.01\% &     1.20\% &    0.02\% \\
HB+MS & SD       &     0.08\% &    0.00\% &     0.04\% &    0.00\% &     0.12\% &    0.00\% &     0.07\% &    0.00\% \\
Other  &         &     0.24\% & $<$0.01\% &     0.17\% &    0.01\% &     0.49\% & $<$0.01\% &     0.39\% &    0.01\% 
\medskip \\
Total  &         &    96.08\% &    3.92\% &    93.58\% &    6.42\% &    98.73\% &    1.27\% &    97.41\% &    2.59\% \\
\hline
\end{tabular} 
\end{table*}

\begin{table*}
\contcaption{}
\begin{tabular}{lccccccccc}
\hline
binary & binary & \multicolumn{2}{c}{model 1} & \multicolumn{2}{c}{model 2} & \multicolumn{2}{c}{model 3} & \multicolumn{2}{c}{model 4} \\
components & type & thin disc & thick disc & thin disc & thick disc & thin disc & thick disc & thin disc & thick disc \\
\hline
\hline
\multicolumn{10}{c}{$n(q) \propto q^{-0.99}$, $0 < q \le 1$} \\
\hline
MS+MS & D        &    78.11\% &    0.31\% &    78.75\% &    0.82\% &    79.53\% &    0.41\% &    80.90\% &    1.07\% \\
HG+MS & D        &     1.36\% &    0.11\% &     1.53\% &    0.25\% &     1.16\% &    0.09\% &     1.23\% &    0.21\% \\
GB+MS & D        &     8.86\% &    1.53\% &     7.86\% &    2.35\% &     3.88\% &    0.29\% &     3.51\% &    0.53\% \\
GB+HG/GB/HB & D  &     0.38\% &    0.06\% &     0.31\% &    0.07\% &     0.74\% &    0.17\% &     0.60\% &    0.20\% \\
HB+MS & D        &     7.08\% &    0.02\% &     5.40\% &    0.02\% &     9.01\% &    0.00\% &     6.57\% &    0.00\% \\
AGB+MS & D       &     0.49\% & $<$0.01\% &     0.35\% & $<$0.01\% &     0.23\% & $<$0.01\% &     0.17\% & $<$0.01\% \\
AGB+HG/GB/HB & D &     0.06\% & $<$0.01\% &     0.04\% & $<$0.01\% &     0.18\% & $<$0.01\% &     0.13\% & $<$0.01\% \\
nHe+MS & D       &     0.08\% &    0.00\% &     0.06\% &    0.00\% &     0.06\% &    0.00\% &     0.04\% &    0.00\% \\
WD+MS & D        &     0.65\% &    0.09\% &     1.31\% &    0.25\% &     1.16\% &    0.14\% &     2.26\% &    0.38\% \\
MS+MS & SD       &     0.30\% & $<$0.01\% &     0.22\% & $<$0.01\% &     1.10\% & $<$0.01\% &     0.82\% & $<$0.01\% \\
HG+MS & SD       &     0.05\% & $<$0.01\% &     0.04\% & $<$0.01\% &     0.19\% & $<$0.01\% &     0.15\% & $<$0.01\% \\
GB+MS & SD       &     0.34\% & $<$0.01\% &     0.26\% & $<$0.01\% &     1.26\% &    0.01\% &     0.94\% &    0.01\% \\
HB+MS & SD       &     0.04\% &    0.00\% &     0.02\% &    0.00\% &     0.09\% &    0.00\% &     0.05\% &    0.00\% \\
Other  &         &     0.09\% & $<$0.01\% &     0.08\% &    0.01\% &     0.30\% & $<$0.01\% &     0.23\% & $<$0.01\% 
\medskip \\
Total  &         &    97.89\% &    2.11\% &    96.23\% &    3.77\% &    98.89\% &    1.11\% &    97.60\% &    2.40\% \\
\hline
\hline
\multicolumn{10}{c}{$M_2$ from same IMF as $M_1$} \\
\hline
MS+MS & D        &    88.17\% &    0.47\% &    86.83\% &    1.16\% &    85.75\% &    0.91\% &    83.93\% &    1.96\% \\
HG+MS & D        &     1.40\% &    0.14\% &     1.53\% &    0.30\% &     1.08\% &    0.13\% &     1.00\% &    0.24\% \\
GB+MS & D        &     5.19\% &    1.10\% &     4.47\% &    1.64\% &     2.87\% &    0.32\% &     2.20\% &    0.48\% \\
GB+HG/GB/HB & D  &     0.15\% &    0.03\% &     0.11\% &    0.04\% &     0.53\% &    0.17\% &     0.35\% &    0.17\% \\
HB+MS & D        &     1.79\% &    0.01\% &     1.22\% &    0.01\% &     2.64\% &    0.00\% &     1.58\% &    0.00\% \\
AGB+MS & D       &     0.11\% & $<$0.01\% &     0.07\% & $<$0.01\% &     0.04\% & $<$0.01\% &     0.03\% & $<$0.01\% \\
AGB+HG/GB/HB & D &     0.01\% & $<$0.01\% &     0.01\% & $<$0.01\% &     0.05\% & $<$0.01\% &     0.03\% & $<$0.01\% \\
nHe+MS & D       &  $<$0.01\% &    0.00\% &  $<$0.01\% &    0.00\% &     0.01\% &    0.00\% &     0.01\% &    0.00\% \\
WD+MS & D        &     1.11\% &    0.17\% &     2.06\% &    0.42\% &     3.93\% &    0.51\% &     6.26\% &    1.09\% \\
MS+MS & SD       &     0.04\% & $<$0.01\% &     0.03\% & $<$0.01\% &     0.30\% & $<$0.01\% &     0.18\% & $<$0.01\% \\
HG+MS & SD       &     0.01\% & $<$0.01\% &     0.01\% & $<$0.01\% &     0.09\% & $<$0.01\% &     0.07\% & $<$0.01\% \\
GB+MS & SD       &     0.08\% & $<$0.01\% &     0.06\% & $<$0.01\% &     0.58\% &    0.01\% &     0.35\% &    0.01\% \\
HB+MS & SD       &  $<$0.01\% &    0.00\% &  $<$0.01\% &    0.00\% &     0.01\% &    0.00\% &     0.01\% &    0.00\% \\
Other  &         &     0.02\% & $<$0.01\% &     0.03\% & $<$0.01\% &     0.06\% &    0.01\% &     0.04\% &    0.01\% 
\medskip \\
Total  &         &    98.08\% &    1.92\% &    96.43\% &    3.57\% &    97.94\% &    2.06\% &    96.04\% &    3.96\% \\
\hline
\end{tabular} 
\end{table*}

In the case of the initial mass ratio distribution $n(q) \propto q$,
the relative contribution of detached double main-sequence star
binaries decreases in favour of detached giant-branch main-sequence
star and horizontal branch main-sequence star binaries. The numbers
for the initial mass ratio distribution $n(q) \propto q^{-0.99}$ show
the opposite effect. The largest shifts in the relative contributions
occur when the secondary mass is assumed to be distributed
independently according to the same IMF as the primary mass. In
particular, the contribution of detached double main-sequence star
binaries then increases to 80-90\%, while the contribution of
detached asymptotic giant-branch main-sequence star binaries decreases
to 5-10\% (see Fig.~\ref{pop}).

\section{Conclusions}

In this exploratory study we used the BiSEPS binary population
synthesis code to estimate the number of eclipsing binaries and single
stars detectable in the Galactic disc by an idealised SuperWASP
exoplanet transit survey. The Galactic disc is modeled by a
superposition of a young (thin) and old (thick) component, both of
which are described by means of a double exponential distribution
function characterised by different scale lengths and scale
heights. The models yield a strong concentration of eclipsing binaries
and single stars towards the Galactic plane and the Galactic centre.

The ratio of the number of detectable eclipsing binaries to the number
of detectable single stars varies by less than a factor of 2.5
across the sky. It is largest in the direction of the Galactic
longitude $l=-7.5^\circ$ and the Galactic latitude $b=-22.5^\circ$, so
that the number of false exoplanet transit detections due to eclipsing
binaries can be expected to be largest in this direction.  More
favourable regions on the sky aimed at minimizing false planetary
detections are therefore located away from this area. The best
conditions are found at Galactic longitudes $75^\circ \la l \la
285^\circ$ and, for Galactic longitudes $|l| \la 75^\circ$, at
Galactic latitudes $|b| \ga 15^\circ$.

Depending on the adopted survey model, the total number of single
stars in the magnitude range detectable by the idealised SuperWASP
survey is of the order of $10^6$--$10^7$. The total number of
detectable eclipsing binaries varies from a few times $10^2$--$10^5$,
with the exact number depending on the adopted survey model (limiting
magnitude and eclipse amplitude) and on the initial mass ratio or
initial secondary mass distribution. In view of the large number of
stars and binaries observed by SuperWASP each night, the lower end of
the predicted number of eclipsing binaries (resulting from the initial
mass ratio distribution $n(q) \propto q^{-0.99}$) could soon be
confronted with interesting and potentially stringent observational
constraints, assuming that our simplifications still capture the
essence of the real SuperWASP set-up.

Provided that complete samples of the most abundant types of binaries
can be compiled by SuperWASP, the relative contribution of each type of
binary to the total population can be used to further constrain the
initial mass ratio or initial secondary mass distribution (see
Fig.~\ref{pop}). In particular, for our standard survey model, the
relative number of double main-sequence star binaries increases from
45\% in the case of the initial mass ratio distribution $n(q)
\propto q$ to 80\% in the case of the initial mass ratio
distribution $n(q) \propto q^{-0.99}$, and 90\% in the case
where the initial secondary mass is distributed independently
according to the same IMF as the primary mass. The contribution of
white dwarf main-sequence star binaries increases from 0.1\% in
the case of the initial mass ratio distribution $n(q) \propto q$ to
0.7\% in the case of the initial mass ratio distribution $n(q)
\propto q^{-0.99}$, and 1.3\% in the case where the initial
secondary mass is distributed independently according to the same IMF
as the primary mass. In contrast, the relative contributions of the
other major groups of binaries tend to decrease when the initial mass
ratio distribution is changed from $n(q) \propto q$ to $n(q) \propto
q^{-0.99}$ or an independent initial secondary mass
distribution. Similar diagnostics for the initial mass ratio or
initial secondary mass distribution can be obtained by considering
ratios of numbers of different types of eclipsing binaries detectable
by the transit survey rather than their fractional contributions to
the total population.

In view of the large variety of eclipsing binaries found, we did not
consider the effects of varying some of the binary evolution
parameters as is customary in binary population synthesis studies
(see, e.g., Willems \& Kolb 2002, 2003, 2004). In particular, the
masses and orbital periods of the eclipsing binaries may be affected
by uncertainties in the treatment of mass transfer if either of the
binary components at some point of its evolution filled its critical
Roche lobe. Additional uncertainties affecting the binary separation
(and thus the possibility of interactions between the component stars
as well as the probability to observe the system as an eclipsing
binary) are introduced by uncertainties in systemic orbital angular
momentum losses through, e.g., magnetic braking and stellar winds. The
latter can furthermore also have a strong impact on the stars
themselves, especially during or after evolutionary phases that are
accompanied by high wind mass-loss rates such as the giant branch,
asymptotic giant branch, and Wolf-Rayet stages of stellar
evolution. The impact of these uncertainties on the relative numbers
and statistical properties of different types of eclipsing binaries
will be addressed in follow-up studies that are tailored more closely
to represent the precise properties of different ongoing and future
exoplanet transit surveys.

The results of our exploratory study reveal very encouraging
differential trends in the collective properties of the eclipsing
binary population that can be probed by extrasolar planet transit
surveys. This suggests that transit surveys will be able to deliver 
stringent constraints on parameters describing the formation and
evolution of binaries once statistically meaningful results from such
surveys are available. 
In future applications of this work we will consider a fine-tuned
transit survey model that more closely matches the actual parameters
of the SuperWASP survey, and that also addresses the fact that the
discovery probability of eclipses will depend on the eclipse
properties. We will also extend our study to other transit surveys.

\section*{Acknowledgements}
We are grateful to Pasi Nurmi and Henri Boffin for providing the
corrected fits to the bolometric corrections given by Flower (1996),
and Carole Haswell, Andy Norton and Sean Ryan for valuable discussions
on the observational aspects of the SuperWASP project. We also thank
the referee, Chris Tout, for a constructive report which helped
improve the paper. Part of this research was supported by the British
Particle Physics and Astronomy Research Council (PPARC). BW
acknowledges the support of NASA ATP grant NAG5-13236 to Vicky
Kalogera and the hospitality of the Open University which allowed the
final stages for this project to be laid out. This research made
extensive use of NASA's Astrophysics Data System Bibliographic
Services.

\bsp

\label{lastpage}


\begin{thebibliography}{}
\bibitem{Al} Alonso R., et al., 2004, ApJ, 613, L153
\bibitem{Ba} Bahcall J.N., 1986, ARA\&A 24, 577
\bibitem{Bi} Binney J., Tremaine S., 1987, Galactic dynamics,
  Princeton University Press 
\bibitem{Bo} Bouchy F., Pont F., Santos N.C., Melo C., Mayor M.,
  Queloz D., Udry S., 2004, A\&A, 421, L13
\bibitem{Br} Brown T.M., 2003, ApJ 593, L125
\bibitem{Ca} Cabrera-Lavers A., Garz{\' o}n F., Hammersley
  P.L. 2005, A\&A, 433, 173  
\bibitem{Ch2} Charbonneau D., 2003, Space Science Reviews, ISSI 
  Workshop on Planetary Systems and Planets in Systems, Eds. S. 
  Udry, W. Benz and R. von Steiger, Dordrecht: Kluwer, in press
\bibitem{Ch} Charbonneau D., Brown T.M., Latham D.W., Mayor M.,
  2000, ApJ 529, L45
\bibitem{Ch3} Charbonneau D., Brown T.M., Dunham E.W., Latham 
  D.W., Looper D.L., Mandushev G., 2003, AIP Conf Proc, 
  The Search for Other Worlds, Eds. S. S. Holt and D. Deming, 
  in press (astro-ph/0401063)
\bibitem{Che} Chen B., et al., 2001, ApJ 553, 184 
\bibitem{Chr} Christian D.J., et al. 2004, Proceedings of the 13th
  Cool Stars Workshop, Ed. F. Favata, in press (astro-ph/0411019) 
\bibitem{Cu} Du C., Zhou X., Ma J., Chen A.B., Yang Y., Li J., 
  Wu H., Jiang Z., Chen J., A\&A 407, 541
\bibitem{Fl} Flower P.J., 1996, ApJ 469, 355
\bibitem{GR} Gilmore G., Reid N., 1983, MNRAS 202, 1025
\bibitem{Ha} Hakkila J., Myers J.M., Stidham B.J., Hartmann D.H.,
  1997, AJ 114, 2042 
\bibitem{He} Henry G.W., Marcy G.W., Butler R.P., Vogt S.S.,
  2000, ApJ 529, L41
\bibitem{Ho} Horne K., 2002, ESA SP-485: Stellar Structure and
  Habitable Planet Finding, 137
\bibitem{Ho2} Horne K., 2003, ASP Conf.~Ser.~294: Scientific Frontiers
  in Research on Extrasolar Planets, 361 
\bibitem{Hu} Hurley J.R., Pols O.R., Tout C.A., 2000, MNRAS
  315, 543
\bibitem{Hu2} Hurley J.R., Tout C.A., Pols O.R., 2002, MNRAS
  329, 897
\bibitem{Ka} Kane S.R., Horne K., Street R.A., Pollacco D.L.,
  James D., Tsapras Y., Collier Cameron A., 2003, ASP 
  Conf.~Ser.~294: Scientific Frontiers in Research on Extrasolar 
  Planets, 387
\bibitem{Ka2} Kane S.R., Collier Cameron A., Horne K., James D.,
Lister T.A., Pollacco D.L., Street R.A., Tsapras Y., 2004,
MNRAS, 353, 689
\bibitem{Ke} Kerber L.O., Javiel S.C., Santiago, B.X., 2001, 
  A\&A 365, 424
\bibitem{Ko} Konacki M., Torres G., Jha S., Sasselov D.D., 2003,
  Nature 421, 507 
\bibitem{Ko2} Konacki M., et al., 2004, ApJ, 609, L37
\bibitem{Ko3} Konacki M., Torres G., Sasselov D.D., Jha S., 2005,
  ApJ, 624, 372  
\bibitem{Kolb} Kolb U., Willems B., 2004, RevMexAA 20, 101
\bibitem{Liu} Liu W.M., Chaboyer B., 2000, ApJ 544, 818 
\bibitem{No} Norris J.E., 1999, Ap\&SS 265, 213 
\bibitem{No2} Norris J.E., 2001, in Encyclopedia of Astronomy and
Astrophysics, Ed. P Murdin, Bristol: IoP Publishing, p 841 
\bibitem{Nu} Nurmi P., Boffin H.M.J., 2003, A\&A 408, 803 
\bibitem{Oj} Ojha D.K., 2001, MNRAS 322, 426
\bibitem{Pol} Pollacco D., 2005, Astronomy and Geophysics, 46, 19 
\bibitem{Po} Pont F., Bouchy F., Queloz D., Santos N.C., Melo C.,
  Mayor M., Udry S., 2004, A\&A, 426, L15
\bibitem{Re} Reid M.J., 1993, ARA\&A 31, 345
\bibitem{Si} Siegel M.H., Majewski S.R., Reid I.N., Thompson
  I.B., 2002, ApJ 578, 151
\bibitem{Sm} Smith L.F., Meynet G., Mermilliod J.-C., 1994, A\&A
  287, 835
\bibitem{St} Street R.A., Pollacco D.L., Fitzsimmons A., Keenan F.P.,
  Horne K., Kane S., Collier Cameron A., Lister T.A., Haswell C.A.,
  Norton A.J., Jones B.W., Skillen I., Hodgkin S., Wheatley P.,
  West R., Brett D., 2003, ASP Conf.~Ser.~294: Scientific Frontiers 
  in Research on Extrasolar Planets, 405 
\bibitem{Wei} Weidemann V., 1990, ARA\&A 28, 103
\bibitem{WK} Willems B., Kolb U., 2002, MNRAS 337, 1004
\bibitem{WK2} Willems B., Kolb, U., 2003, MNRAS 343, 949
\bibitem{WK3} Willems B., Kolb U., 2004, A\&A 419, 1057
\end{thebibliography}
\end{document}